\documentclass[conference]{IEEEtran}

\usepackage{cite}
\usepackage{amsmath}
\usepackage{graphicx}
\usepackage{hyperref}
\usepackage{algorithm}
\usepackage{algorithmic}
\usepackage{float}

\title{Towards Post-Quantum Secure Pharmacovigilance with ML-KEM and ML-DSA}

\author{
\IEEEauthorblockN{
Saee Desai$^{*}$,
Tom Shimoni,
Eddie Cameron,
David Akamine,
Aniketh Chunduri
}

\IEEEauthorblockA{
\small
Do Quantum, University of Maryland\\
College Park, Maryland, USA\\
sdesai09@umd.edu, tshimoni@terpmail.umd.edu,\\
ecamero1@terpmail.umd.edu, dakamine@terpmail.umd.edu,\\
achundur@terpmail.umd.edu\\
\vspace{0.3em}
$^{*}$Corresponding author
}
}

\begin{document}

\maketitle

% ================= ABSTRACT =================
\begin{abstract}

Pharmacovigilance systems handle sensitive healthcare and drug-safety data, including adverse event reports and clinical observations. As quantum computing advances, classical public-key cryptographic systems such as RSA and elliptic-curve cryptography may become vulnerable, creating long-term risks for healthcare data that must remain confidential for many years.

This paper presents an educational prototype of a post-quantum secure pharmacovigilance data pipeline. The system uses ML-KEM-768 for post-quantum key establishment, HKDF-SHA-256 for deriving an AES key, AES-256-GCM for efficient file encryption, and ML-DSA-65 for digital signatures and tamper detection. The pipeline supports multiple file formats, including TXT, CSV, JSON, and PDF, by treating files as raw bytes and preserving metadata for reconstruction at the receiver.

The prototype includes separate hospital, gateway, pharma receiver, attacker, benchmarking, and dashboard components. We evaluate the system using synthetic pharmacovigilance datasets of different sizes and formats. Our results show that ML-KEM adds a small constant overhead, while AES encryption and ML-DSA signing dominate runtime as file size increases. This work is not a production-ready healthcare system, but rather an educational systems-level exploration of how post-quantum cryptographic primitives can be integrated into healthcare-style data pipelines.

\end{abstract}

% ================= INTRODUCTION =================
\section{Introduction}

Pharmacovigilance systems are responsible for monitoring the safety of drugs after they are released to the market~\cite{who_pharmacovigilance}. These systems collect and process sensitive healthcare data, including adverse drug reactions, patient reports, and clinical observations. Protecting this data is critical, as it often contains personally identifiable and long-term medical information.

Currently, most public-key secure communication systems rely on classical cryptographic techniques such as RSA and elliptic-curve cryptography (ECC)~\cite{nist_pqc_migration}. However, the development of quantum computing poses a potential threat to these systems. Quantum algorithms, such as Shor’s algorithm, could threaten widely used public-key cryptographic schemes~\cite{shor1997}, making encrypted data vulnerable in the future.

A key concern is the “harvest now, decrypt later” (HNDL) threat model~\cite{nist_pqc_migration}. In this scenario, adversaries may collect encrypted healthcare data today and store it with the intention of decrypting it once quantum computers become powerful enough. Since pharmacovigilance data often needs to remain confidential for many years, this creates a long-term security risk.

Post-quantum cryptography (PQC) aims to address this problem by introducing cryptographic algorithms that are believed to be secure against both classical and quantum attacks. Recently, NIST standardized several lattice-based PQC algorithms, including ML-KEM (Module-Lattice-Based Key Encapsulation Mechanism)~\cite{fips203} for key encapsulation and ML-DSA (Module-Lattice-Based Digital Signature Algorithm)~\cite{fips204} for digital signatures.

In this work, we explore how post-quantum cryptography can be applied to a pharmacovigilance data pipeline. Instead of implementing the algorithms from scratch, we focus on integrating existing PQC implementations into a realistic data flow involving a hospital, a gateway, and a pharmaceutical analysis system.

We design and implement an end-to-end prototype that:
\begin{itemize}
    \item Uses ML-KEM for secure key establishment
    \item Uses HKDF to derive symmetric keys
    \item Uses AES-256-GCM for efficient data encryption
    \item Uses ML-DSA for data integrity and authenticity
\end{itemize}

We also evaluate the system using synthetic pharmacovigilance datasets of different sizes and formats, and simulate attack scenarios to demonstrate how tampered data is detected.

This work serves as an educational and practical exploration of how PQC can be integrated into real-world healthcare data pipelines, highlighting both its feasibility and its current limitations.

% ================= MOTIVATION =================
\section{Motivation}

Pharmacovigilance is defined as the science and activities related to the detection, assessment, understanding, and prevention of adverse drug effects~\cite{who_pharmacovigilance}. These systems play a critical role in ensuring drug safety and protecting public health~\cite{adr_importance}. They rely on continuous data collection from hospitals, clinical trials, and healthcare providers, including adverse drug reactions and patient observations.

A key characteristic of pharmacovigilance systems is the long-term value of the data they handle. Adverse event reports, patient histories, and clinical observations are often stored and analyzed over extended periods of time. This makes the confidentiality of such data not just a short-term concern, but a long-term requirement.

Since pharmacovigilance data must often remain confidential for many years, it becomes particularly vulnerable under future cryptographic threats. Even if current encryption mechanisms are secure today, they may not remain secure in the presence of future advances in quantum computing.

This concern is captured by the “harvest now, decrypt later” (HNDL) threat model~\cite{nist_pqc_migration}. In this scenario, adversaries collect encrypted healthcare data today with the intention of decrypting it in the future when quantum capabilities improve. For pharmacovigilance systems, this is especially critical because historical patient data and adverse event reports could be exposed long after they were originally transmitted.

In addition to confidentiality, pharmacovigilance pipelines must ensure data integrity and authenticity. Data exchanged between hospitals, pharmaceutical companies, and regulatory bodies must not be altered in transit, and the origin of the data must be verifiable. Any tampering with adverse event reports or clinical observations could lead to incorrect conclusions about drug safety, potentially affecting clinical decisions and patient outcomes.

Post-quantum cryptography (PQC) offers a promising approach to addressing these challenges. By using cryptographic algorithms that are designed to be secure against both classical and quantum attacks, it becomes possible to protect healthcare data against future threats. In particular, ML-KEM enables secure key establishment, while ML-DSA provides mechanisms for ensuring data authenticity and integrity.

The motivation of this work is not to develop new cryptographic algorithms, but to understand how existing PQC primitives can be integrated into a realistic pharmacovigilance data pipeline. By building and evaluating an end-to-end system, we aim to demonstrate how post-quantum security can be applied in practice to protect sensitive healthcare data.

% ================= PLACEHOLDERS =================
\section{Related Work}

The emergence of quantum computing has led to significant research in post-quantum cryptography (PQC), which aims to develop cryptographic algorithms that are secure against both classical and quantum adversaries. The National Institute of Standards and Technology (NIST) has recently standardized several PQC algorithms, including ML-KEM for key encapsulation and ML-DSA for digital signatures~\cite{fips203, fips204}. These algorithms are based on lattice-based cryptography and are considered strong candidates for replacing classical public-key systems such as RSA and elliptic-curve cryptography.

Beyond theoretical development, there has been growing interest in integrating PQC into real-world systems. Several studies and industry efforts have explored the use of PQC in secure communication protocols, particularly Transport Layer Security (TLS). Hybrid approaches combining classical and post-quantum algorithms have been explored in real-world systems to ensure backward compatibility while providing quantum resistance~\cite{pqc_tls}. These efforts highlight the practical challenges of adopting PQC, including performance overhead and compatibility with existing infrastructure. In addition to protocol-level integration challenges, there is growing interest in understanding how standardized PQC primitives can be incorporated into domain-specific systems that require long-term confidentiality and integrity guarantees.

Another area of research focuses on evaluating the computational and communication costs associated with PQC algorithms. Compared to classical cryptographic methods, PQC schemes often involve larger key sizes and higher computational overhead~\cite{kyber_spec}. These factors can impact latency and scalability, especially in systems that handle large volumes of data or operate under resource constraints.

In the context of healthcare and data-sensitive applications, security and privacy have been widely studied, particularly in areas such as electronic health records and clinical data exchange~\cite{who_pharmacovigilance}. However, relatively limited work has explored the integration of post-quantum cryptographic primitives into pharmacovigilance pipelines, where long-term data confidentiality and integrity are critical.

This work addresses this gap by designing and implementing an end-to-end pharmacovigilance data pipeline that incorporates ML-KEM, ML-DSA, and symmetric encryption. Unlike prior work that focuses either on cryptographic primitives or protocol-level integration, our approach emphasizes system-level design and practical evaluation using realistic data formats and attack scenarios.

\section{System Architecture}

The proposed system models a post-quantum secure pharmacovigilance data pipeline involving multiple entities responsible for generating, transmitting, verifying, and analyzing healthcare data. The architecture is designed to reflect a simplified but realistic data flow between a hospital, a gateway server, and a pharmaceutical analysis system.

\subsection{System Overview}

The system consists of four primary components:

\begin{itemize}
    \item \textbf{Hospital:} Generates pharmacovigilance data and initiates secure transmission.
    \item \textbf{Gateway Server:} Acts as an intermediate node that performs key decapsulation and logs incoming data.
    \item \textbf{Pharmaceutical System:} Verifies data authenticity and decrypts the received information.
    \item \textbf{Attacker Module:} Simulates adversarial behavior by modifying transmitted data to test system robustness.
\end{itemize}
\begin{figure}[h]
\centering
\includegraphics[width=0.5\textwidth]{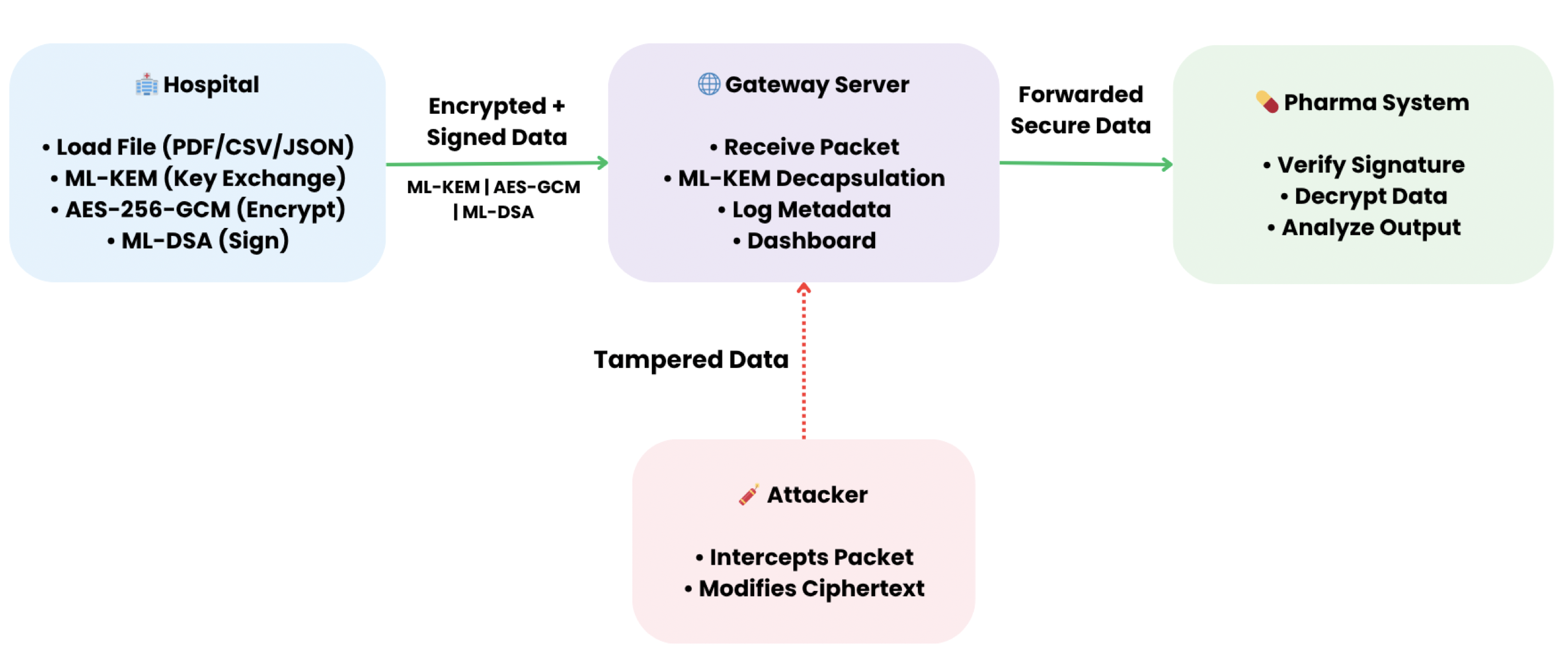}
\caption{Post-Quantum Pharmacovigilance Data Pipeline. The hospital encrypts and signs data before sending it to the gateway, which forwards it to the pharmaceutical system. An attacker may attempt to tamper with the data in transit.}
\label{fig:systemoverview}
\end{figure}
The overall data flow follows a sender-receiver model, where the hospital securely transmits data to the pharmaceutical system through the gateway.

\subsection{Secure Data Transmission Pipeline}

The pipeline integrates post-quantum and symmetric cryptographic primitives in a layered manner:

\begin{enumerate}
    \item \textbf{Key Establishment (ML-KEM):}  
    The hospital uses the public key of the gateway to perform ML-KEM encapsulation, generating a shared secret and a corresponding ciphertext. The gateway later uses its private key to decapsulate this ciphertext and recover the same shared secret.

    \item \textbf{Key Derivation (HKDF):}  
    The shared secret produced by ML-KEM is passed through a key derivation function (HKDF with SHA-256) to produce a 256-bit symmetric encryption key.

    \item \textbf{Data Encryption (AES-256-GCM):}  
    The hospital encrypts pharmacovigilance data using AES-256 in Galois/Counter Mode (GCM), which provides both confidentiality and integrity at the encryption level.

    \item \textbf{Digital Signature (ML-DSA):}  
    The encrypted payload is signed using ML-DSA to ensure authenticity and detect any tampering during transmission.

    \item \textbf{Data Transmission:}  
    The hospital sends a structured payload containing the ML-KEM ciphertext, AES-encrypted data, signature, and metadata to the gateway server.

\end{enumerate}

\subsection{Data Verification and Decryption}

On the receiver side, the pharmaceutical system performs the following steps:

\begin{enumerate}
    \item \textbf{Signature Verification:}  
    The received ciphertext is verified using the hospital's public key. If verification fails, the data is rejected.

    \item \textbf{Key Recovery:}  
    The gateway decapsulates the ML-KEM ciphertext to recover the shared secret, which is then used to derive the AES key.

    \item \textbf{Data Decryption:}  
    If the signature is valid, the pharmaceutical system decrypts the data using AES-256-GCM to obtain the original pharmacovigilance dataset.
\end{enumerate}

\subsection{Attack Simulation}

To evaluate system robustness, an attacker module intercepts and modifies transmitted ciphertext before resending it to the gateway. Since the signature corresponds to the original ciphertext, any modification results in a verification failure at the receiver, demonstrating the effectiveness of ML-DSA in detecting tampering.

\subsection{Supported Data Formats}

The system is designed to handle multiple data formats, including TXT, CSV, JSON, and PDF. Instead of processing structured data directly, all files are treated as raw byte streams during encryption and decryption. Metadata such as file type and size is preserved to enable correct reconstruction at the receiver.

\subsection{System Visualization}

A lightweight web-based dashboard is implemented at the gateway to visualize system activity. It displays incoming packets, verification status, detected attacks, and benchmark metrics, providing an intuitive representation of the secure data pipeline.

\section{Implementation}

The proposed pharmacovigilance pipeline is implemented as a modular system in Python, with separate components representing the hospital, gateway server, pharmaceutical receiver, attacker, benchmarking module, and visualization dashboard. The implementation focuses on integrating post-quantum cryptographic primitives into a realistic data flow rather than developing the cryptographic algorithms from scratch.

\subsection{Cryptographic Primitives}

The system uses ML-KEM for key encapsulation and ML-DSA for digital signatures, based on the NIST-standardized specifications~\cite{fips203, fips204}. According to NIST security categories, ML-KEM-768 and ML-DSA-65 correspond to Category 3 security levels, intended to provide security strength comparable to AES-192 against classical adversaries~\cite{nist_ir_8547}. Under quantum attack models, symmetric-key security levels are reduced due to Grover-style search speedups. Instead of implementing these algorithms manually, we use existing open-source Python implementations by Giacomo Pope~\cite{pope_kyber_py, pope_dilithium_py}, which provide educational implementations of ML-KEM and ML-DSA.

For symmetric encryption, AES-256 in Galois/Counter Mode (GCM) is used to ensure both confidentiality and integrity of the data. A 256-bit encryption key is derived from the ML-KEM shared secret using the HKDF key derivation function with SHA-256.

\subsection{System Components}

The system is divided into the following modules:

\begin{itemize}
    \item \textbf{Hospital Module:}  
    Responsible for loading input files, generating metadata, performing ML-KEM encapsulation, deriving the AES key, encrypting data using AES-GCM, and signing the encrypted payload using ML-DSA. The module supports multiple file formats and treats all inputs as raw byte streams.

    \item \textbf{Gateway Server:}  
    Implemented using a lightweight Flask server, the gateway receives incoming packets, performs ML-KEM decapsulation to recover the shared secret, and logs the received data. For demonstration purposes, the recovered session key is stored in logs, although this would not be done in a real deployment.

    \item \textbf{Pharmaceutical Receiver:}  
    Retrieves packets from the gateway, verifies the ML-DSA signature, and decrypts the ciphertext using the derived AES key. If signature verification fails, the data is rejected.

    \item \textbf{Attacker Module:}  
    Simulates adversarial behavior by intercepting and modifying ciphertext before resending it to the gateway. This module is used to demonstrate that tampered data fails signature verification.

    \item \textbf{Benchmarking Module:}  
    Measures the performance of the pipeline by evaluating key operations such as ML-KEM encapsulation and decapsulation, AES encryption and decryption, and ML-DSA signing and verification across different data sizes and formats.

    \item \textbf{Dashboard Interface:}  
    A web-based interface integrated into the gateway server displays real-time logs, packet metadata, verification status, and benchmark results, providing a visual representation of system activity.
\end{itemize}

\subsection{File Handling and Data Formats}

To simulate realistic pharmacovigilance data, synthetic datasets are generated in multiple formats, including TXT, CSV, JSON, and PDF. All files are processed as raw byte streams during encryption and decryption, ensuring format independence at the cryptographic layer. Metadata such as filename, file type, and data size is included in the transmitted payload to enable correct reconstruction at the receiver.

\subsection{Data Flow and Payload Structure}

The hospital transmits a structured payload to the gateway containing the following elements:

\begin{itemize}
    \item ML-KEM ciphertext (for key recovery)
    \item AES-encrypted data (nonce + tag + ciphertext)
    \item ML-DSA signature
    \item Metadata (source, timestamp, file information)
\end{itemize}

\begin{algorithm}[h]
\caption{Secure Data Transmission Pipeline}
\begin{algorithmic}[1]
\REQUIRE Input file $F$, pre-shared public/private keys $(pk, sk)$
\ENSURE Verified and decrypted data at receiver

\STATE \textbf{// Sender Side (Hospital)}

\STATE Generate shared secret using ML-KEM:
\STATE $(ss, ct_{kem}) \leftarrow \text{Encaps}(pk)$

\STATE Derive AES key using HKDF:
\STATE $k_{aes} \leftarrow \text{HKDF}(ss)$

\STATE Encrypt file using AES-GCM:
\STATE $ct_{aes} \leftarrow \text{Encrypt}(k_{aes}, F)$

\STATE Sign ciphertext using ML-DSA:
\STATE $\sigma \leftarrow \text{Sign}(sk, ct_{aes})$

\STATE Send $(ct_{kem}, ct_{aes}, \sigma)$ to receiver

\vspace{0.3em}
\STATE \textbf{// Receiver Side (Gateway / Pharma System)}

\STATE Recover shared secret:
\STATE $ss \leftarrow \text{Decaps}(ct_{kem})$

\STATE Derive AES key:
\STATE $k_{aes} \leftarrow \text{HKDF}(ss)$

\STATE Verify signature:
\IF{Verify($ct_{aes}, \sigma$) = false}
    \STATE Reject data
\ENDIF

\STATE Decrypt ciphertext:
\STATE $F \leftarrow \text{Decrypt}(k_{aes}, ct_{aes})$

\RETURN $F$
\end{algorithmic}
\end{algorithm}

This structure ensures that the receiver has all necessary components to verify and decrypt the data securely.

\subsection{Experimental Setup}

Experiments are conducted using synthetic pharmacovigilance datasets of varying sizes (e.g., 1 MB, 10 MB, and larger). The benchmarking module records execution times for cryptographic operations and evaluates system performance across different data formats. All experiments are performed on a standard computing environment without hardware acceleration.

\section{Attack Model and Security Analysis}

\subsection{Threat Model}

The system considers an adversarial setting in which an attacker can intercept, modify, and resend data transmitted between the hospital and the gateway. This models a realistic network-level attacker with the capability to observe and tamper with in-transit data, but without access to private cryptographic keys.

We assume that the attacker does not compromise the internal systems of the hospital or the pharmaceutical receiver, and does not possess the private keys required for ML-KEM decapsulation or ML-DSA signing.

\subsection{Attack Scenarios}

To evaluate the robustness of the proposed pipeline, we simulate multiple attack scenarios targeting the integrity and authenticity of transmitted data.

\begin{itemize}
    \item \textbf{Ciphertext Tampering Attack:}  
    The attacker modifies one or more bytes in the encrypted payload before resending it to the gateway. Since the digital signature corresponds to the original ciphertext, any modification results in a mismatch during signature verification.

    \item \textbf{Random Corruption Attack:}  
    The attacker introduces random changes at multiple positions in the ciphertext, simulating noisy or malicious alterations. This also leads to signature verification failure.

    \item \textbf{Truncation Attack:}  
    The attacker removes a portion of the ciphertext before forwarding it. This results in decryption failure due to authentication tag mismatch in AES-GCM.

    \item \textbf{Replay-Based Modification:}  
    The attacker modifies the ciphertext while reusing the original signature. Since the signature is bound to the original message, verification fails at the receiver.

    \item \textbf{Metadata Manipulation (Limitation Case):}  
    The attacker modifies metadata fields such as source or file size without altering the ciphertext. Since the current implementation signs only the encrypted payload, such modifications may not be detected. This highlights a limitation of the current design.
\end{itemize}

\subsection{Security Properties}

The system achieves the following security properties:

\begin{itemize}
    \item \textbf{Confidentiality:}  
    Data is encrypted using AES-256-GCM, with keys derived from ML-KEM shared secrets. Without access to the shared secret, an attacker cannot decrypt the data.

    \item \textbf{Integrity:}  
    Any modification to the ciphertext is detected through ML-DSA signature verification or AES-GCM authentication failure.

    \item \textbf{Authenticity:}  
    The use of ML-DSA ensures that only data signed by the legitimate sender is accepted by the receiver.

    \item \textbf{Post-Quantum Security:}  
    The use of ML-KEM and ML-DSA is designed to provide resistance against quantum adversaries, assuming the hardness of underlying lattice-based problems.
\end{itemize}

\subsection{Discussion}

The attack simulations demonstrate that the system demonstrates the detection of tampering attempts on encrypted data. In particular, ciphertext modifications consistently result in signature verification failure, preventing corrupted data from being processed by the pharmaceutical system.

However, the current design signs only the encrypted payload and not the associated metadata. As a result, metadata-only manipulation may go undetected. This limitation suggests that future implementations should incorporate authenticated metadata or include metadata within the signed message to provide complete integrity protection.

Overall, the results show that integrating ML-DSA with authenticated encryption provides a meaningful defense against common network-level attacks in a pharmacovigilance data pipeline.

\section{Results and Evaluation}

\subsection{Experimental Setup}

The system is evaluated using synthetic pharmacovigilance datasets of varying sizes and formats, including TXT, CSV, JSON, and PDF files. Data sizes range from approximately 1 MB to 10 MB to analyze scalability and performance behavior.

All experiments are conducted using a Python-based implementation without hardware acceleration. The benchmarking module measures execution time for each stage of the pipeline, including ML-KEM encapsulation and decapsulation, AES encryption and decryption, and ML-DSA signing and verification. Each experiment is repeated multiple times, and average values are reported.

\subsection{Performance Metrics}

The following metrics are used to evaluate system performance:

\begin{itemize}
    \item \textbf{KEM Time:} Time required for ML-KEM encapsulation and decapsulation
    \item \textbf{Encryption Time:} Time required for AES-256-GCM encryption
    \item \textbf{Signing Time:} Time required for ML-DSA signature generation
    \item \textbf{Verification Time:} Time required for ML-DSA signature verification
    \item \textbf{Decryption Time:} Time required for AES-GCM decryption
    \item \textbf{Total Time:} End-to-end pipeline execution time
\end{itemize}

\subsection{Performance Results}

Table~\ref{tab:results} summarizes the performance of the system for representative dataset sizes.

\begin{table}[H]
\centering
\caption{End-to-End Performance Across File Formats}
\label{tab:results}
\begin{tabular}{|c|c|c|c|}
\hline
Format & Size & Total (ms) & Sign (ms) \\
\hline
TXT & 1 MB & 89.22 & 47.88 \\
TXT & 10 MB & 281.01 & 60.11 \\
CSV & 1.02 MB & 86.15 & 49.28 \\
CSV & 10.22 MB & 277.00 & 56.67 \\
JSON & 1 MB & 77.41 & 41.13 \\
JSON & 10 MB & 311.04 & 93.07 \\
PDF & 1 MB & 79.33 & 42.13 \\
PDF & 10 MB & 298.46 & 80.58 \\
\hline
\end{tabular}
\end{table}

From the table, it can be observed that total processing time increases significantly as data size grows from approximately 1 MB to 10 MB. This trend is consistent across all file formats, indicating that system performance is primarily influenced by payload size rather than file structure.

\begin{figure}[H]
\centering
\includegraphics[width=0.48\textwidth]{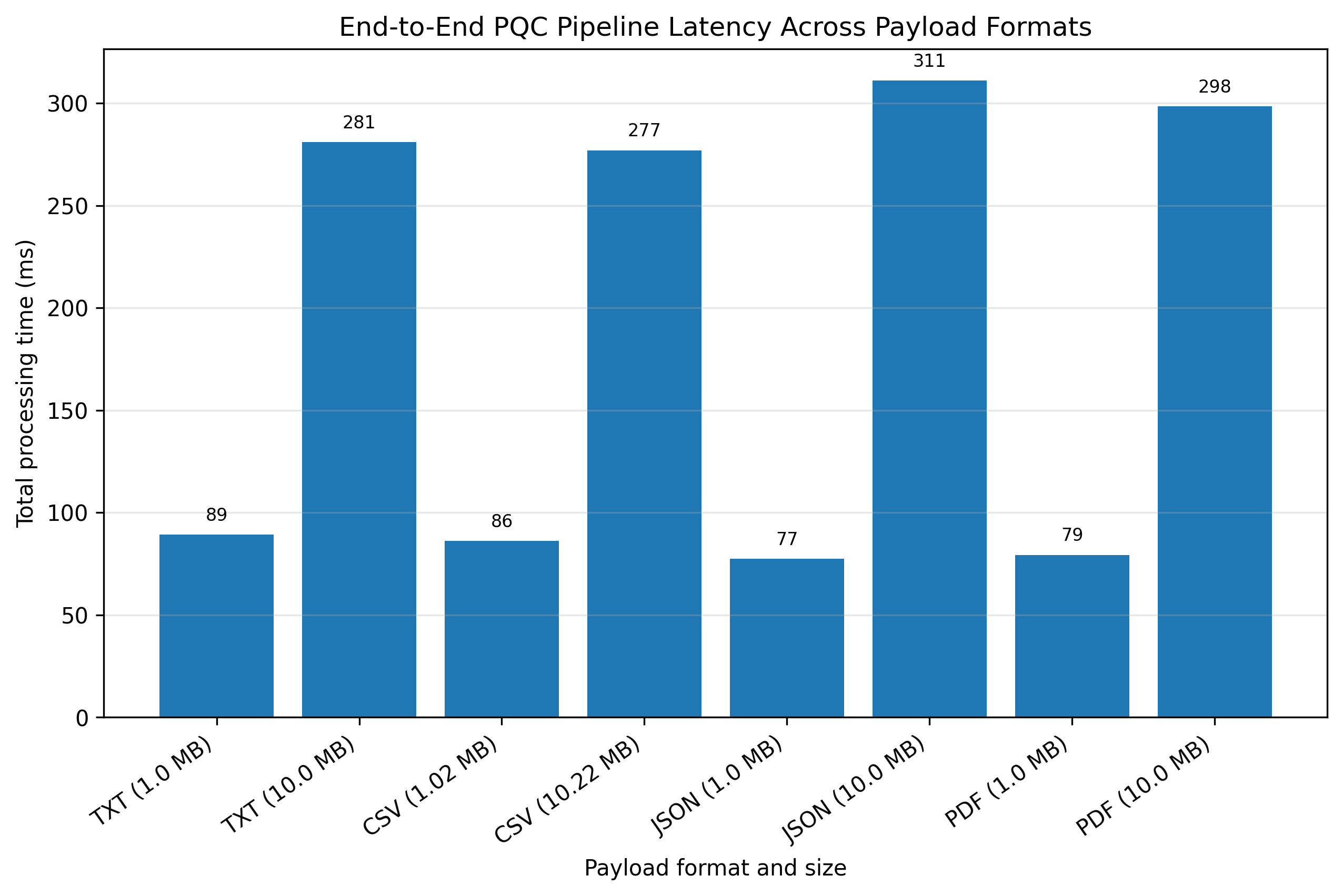}
\caption{End-to-end processing time across payload formats and sizes.}
\label{fig:total_latency}
\end{figure}

As shown in Fig.~\ref{fig:total_latency}, total processing time increases with payload size across all file formats.

The figure further highlights this behavior by showing a clear upward trend in total latency with increasing data size. While minor variations exist between formats, these differences are relatively small compared to the overall impact of data size. This confirms that the pipeline processes all file types uniformly by operating on raw byte streams.

\subsection{Analysis}

\begin{figure}[H]
\centering
\includegraphics[width=0.48\textwidth]{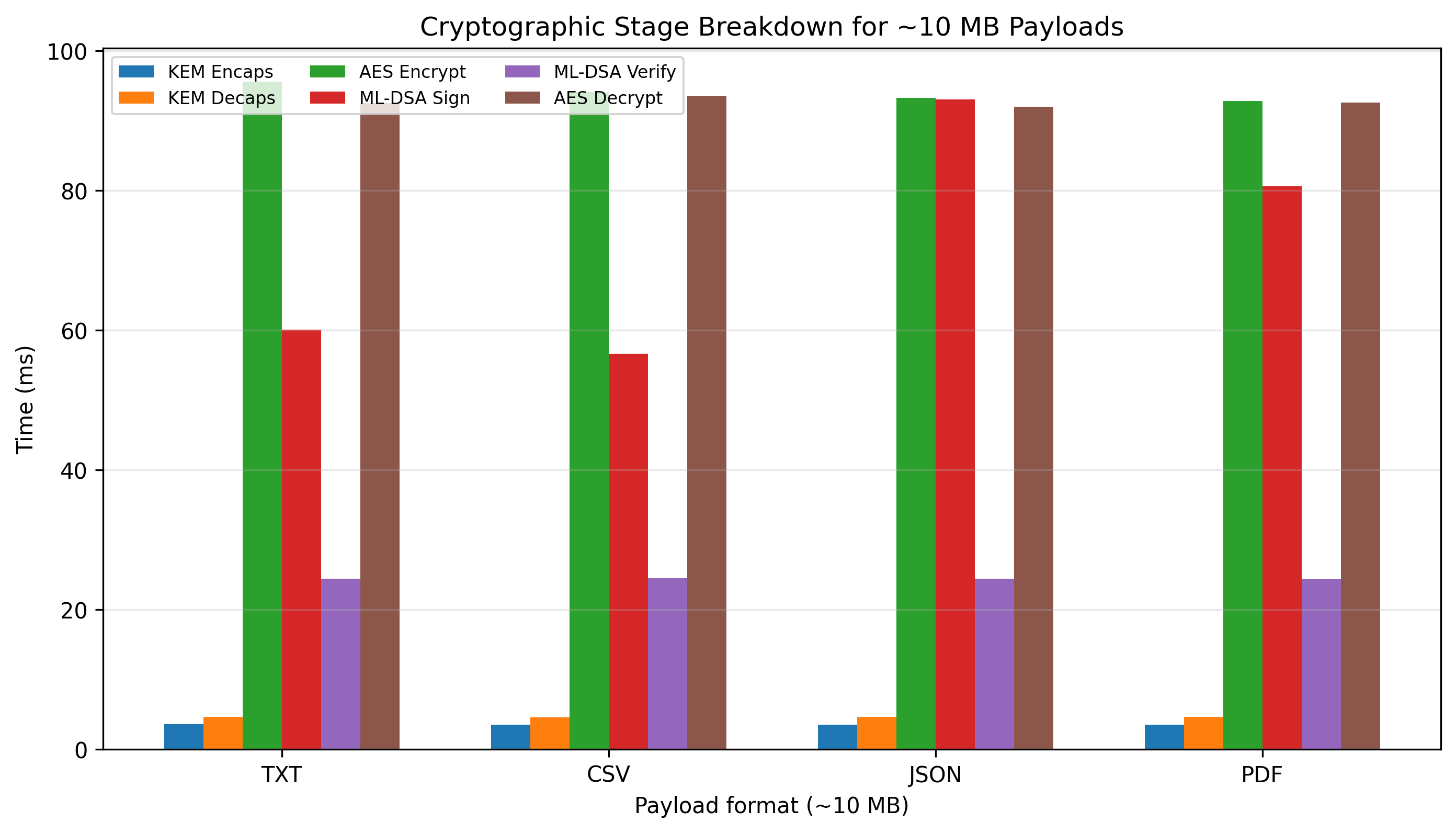}
\caption{Cryptographic stage breakdown for approximately 10 MB payloads.}
\label{fig:stage_breakdown}
\end{figure}

As shown in Fig.~\ref{fig:stage_breakdown}, AES operations and ML-DSA signing contribute the largest portion of runtime for larger payloads.

The results reveal several key trends:

\begin{itemize}
    \item \textbf{Constant KEM Overhead:}  
    The time required for ML-KEM encapsulation and decapsulation remains nearly constant across all data sizes. This is expected, as key exchange operates independently of payload size.

    \item \textbf{Linear Encryption Scaling:}  
    AES encryption and decryption times increase linearly with data size. This indicates that the primary performance cost for large datasets comes from symmetric encryption rather than post-quantum operations.

    \item \textbf{Signature Cost:}  
    ML-DSA signing time is higher than verification time and contributes a noticeable portion of total latency, particularly for smaller datasets.

    \item \textbf{Format Independence:}  
    The system demonstrates consistent performance across TXT, CSV, JSON, and PDF formats. Since all data is processed as raw bytes, file format does not significantly impact cryptographic performance.

\end{itemize}

\subsection{Scalability Observations}

As data size increases from 1 MB to 10 MB, total processing time increases proportionally, primarily due to AES encryption and decryption. In contrast, ML-KEM overhead remains negligible in comparison, indicating that post-quantum key establishment does not significantly impact scalability.

These observations suggest that the proposed pipeline can handle larger pharmacovigilance datasets efficiently, with performance primarily dependent on symmetric encryption throughput.

\subsection{Discussion}

The evaluation demonstrates that integrating post-quantum cryptographic primitives into a pharmacovigilance pipeline is computationally feasible. While ML-DSA introduces additional overhead compared to classical signatures, the overall system remains practical for moderate data sizes.

The results also highlight the importance of combining post-quantum key establishment and authentication with symmetric encryption for efficient bulk data protection.

\section{Limitations}

While the proposed system demonstrates the feasibility of integrating post-quantum cryptographic primitives into a pharmacovigilance data pipeline, several limitations must be acknowledged.

\subsection{Educational Implementation}

The system uses existing open-source Python implementations of ML-KEM and ML-DSA~\cite{pope_kyber_py, pope_dilithium_py}, which are intended for educational purposes. These implementations are not optimized for production use and have not undergone formal security validation or compliance testing.

\subsection{Simplified System Model}

The architecture models a simplified pharmacovigilance pipeline with clearly defined roles (hospital, gateway, pharmaceutical system). Real-world healthcare systems are significantly more complex and involve additional components such as secure communication protocols (e.g., TLS), access control mechanisms, and integration with electronic health record (EHR) systems.

\subsection{Session Key Exposure (Demonstration Artifact)}

For demonstration and debugging purposes, the gateway logs the recovered shared secret derived from ML-KEM. In a real-world deployment, exposing session keys in this manner would introduce serious security risks and would not be acceptable.

\subsection{Metadata Integrity Limitation}

In the current design, only the encrypted payload is signed using ML-DSA. Metadata fields such as file name, source, and timestamp are not included in the signature. As a result, metadata-only modifications may not be detected by the system.

\subsection{Performance Environment}

All experiments are conducted in a software-only environment without hardware acceleration. Performance results may differ in real-world deployments that utilize optimized cryptographic libraries, hardware security modules, or dedicated accelerators.

\subsection{Scope of Security Evaluation}

The attack model focuses primarily on network-level attacks, such as ciphertext tampering and replay-style modifications. The system does not address more advanced threat scenarios, including key compromise, insider attacks, or side-channel vulnerabilities.

\section{Conclusion}

This work presents an end-to-end prototype of a post-quantum secure pharmacovigilance data pipeline. By integrating ML-KEM for key establishment, HKDF for key derivation, AES-256-GCM for data encryption, and ML-DSA for digital signatures, the system explores how post-quantum cryptographic primitives can be integrated in a realistic healthcare data flow.

The implementation models a simplified but representative interaction between a hospital, a gateway server, and a pharmaceutical analysis system. Through attack simulations, the system shows that ciphertext tampering and modification attempts are effectively detected using digital signatures and authenticated encryption.

Performance evaluation across multiple data sizes and file formats indicates that the primary computational overhead arises from symmetric encryption and digital signatures, while the cost of post-quantum key establishment remains relatively small and constant. These results suggest that integrating post-quantum cryptography into pharmacovigilance pipelines is computationally feasible for moderate-scale data exchange.

Overall, this work highlights the practicality of adopting post-quantum cryptographic techniques in healthcare data systems, while emphasizing the importance of system-level design and integration.

\section{Future Work}

While the current prototype demonstrates the feasibility of a post-quantum secure pharmacovigilance pipeline, several directions remain for future improvement and exploration.

First, future work can extend the system to integrate post-quantum cryptographic primitives into standard communication protocols such as TLS, enabling secure end-to-end communication in real-world deployments. Hybrid approaches combining classical and post-quantum algorithms can also be explored to ensure backward compatibility during the transition phase.

Second, metadata integrity can be improved by incorporating metadata into the signed payload or by using authenticated data structures, ensuring complete protection against tampering.

Third, performance can be optimized using production-grade cryptographic libraries and hardware acceleration, allowing evaluation at larger data scales and under realistic system constraints.

Finally, future research can explore more advanced threat models, including insider attacks, key compromise scenarios, and side-channel vulnerabilities, to provide a more comprehensive security evaluation.

\section*{Acknowledgements}

This work was conducted as part of the Do Quantum student organization at the University of Maryland.

We acknowledge Giacomo Pope for providing open-source implementations of ML-KEM and ML-DSA, which were used as part of this educational prototype.

We thank all team members for their collaboration in designing, implementing, and evaluating the system.

\bibliographystyle{IEEEtran}
\bibliography{references}

\appendices

\section{Project Resources}

The complete implementation of the proposed system is available at:

\begin{itemize}
    \item GitHub Repository: \url{https://github.com/saeedesai-coder/Pharma-PQC}
\end{itemize}

\section{Data Availability}

The datasets analyzed in this work were synthetically generated using custom Python scripts developed as part of the project workflow. The scripts used for dataset generation, along with the implementation code, are available in the project repository: \url{https://github.com/saeedesai-coder/Pharma-PQC}

Additional generated datasets may be requested from the corresponding author.

\typeout{get arXiv to do 4 passes: Label(s) may have changed. Rerun}
\end{document}